\documentclass{article}
\usepackage{times}  % Use Times font (recommended for arXiv)
\usepackage{lmodern}  % Use Latin Modern, an improved version of Computer Modern
\usepackage{courier}  % Use Courier for monospaced text (code snippets)
\usepackage{graphicx}  % For including figures
\usepackage{hyperref}  % For clickable links
\usepackage{amsmath, amssymb}  % For mathematical symbols
\usepackage{algorithm, algorithmic}  % For pseudocode (if needed)
\usepackage{float}

\title{Bridging LLM-Generated Code and Requirements: Reverse Generation technique and SBC Metric for Developer Insights}
\author{Ahilan Ayyachamy Nadar Ponnusamy\\ \texttt{ahilanp@gmail.com}}
\date{\today}

\begin{document}

\maketitle

\begin{abstract}
The rise of Large Language Models (LLMs) in software engineering, particularly in code generation, has garnered significant attention. However, assessing the quality of AI-generated code remains a challenge due to the inherent complexity of programming tasks and the lack of robust evaluation metrics that align well with human judgment. Traditional token-based metrics such as BLEU and ROUGE, while commonly used in natural language processing, exhibit weak correlations with human assessments in code intelligence and verification tasks. Furthermore, these metrics are primarily research focused and are not designed for seamless integration into the software development lifecycle, limiting their practical utility for developers seeking to improve code quality and security.

AI-assisted coding has been shown to be more beneficial for senior developers, as they possess the expertise to critically evaluate the generated code for correctness, completeness, and compliance. In contrast, junior developers may struggle to identify hallucinations, missing functionality, or incorrect logic in AI-generated code. To bridge this gap, This paper introduces a novel scoring mechanism called the \textbf{SBC score}, which is based on a reverse generation technique that leverages the natural language generation capabilities of LLMs. Unlike direct code analysis, our approach reconstructs system requirements from AI-generated code and compares them with the original specifications to quantify accuracy. The SBC score combines \textbf{semantic similarity, BLEU, and completeness analysis}, providing actionable insights to developers by highlighting missing features and hallucinations. This hybrid metric not only improves the evaluation of AI-generated code but also offers a real-time, interpretable scoring system that can be integrated into the software development process, benefiting developers of all experience levels.  \textbf{Our code and datasets are available on GitHub: }\href{https://github.com/AhilanPonnusamy/Reverse-Generation-and-SBC-Metric--Reference-Implementation}{GitHub Repository}.

\end{abstract}

\section{Introduction}

AI-powered code assistants, leveraging the power of Large Language Models (LLMs), are becoming a focal point for enterprises, offering promising capabilities in automating code generation. However, evaluating the quality of LLM-generated code remains a complex challenge due to the intricacies of programming concepts and syntax, which differ significantly from natural language generation \cite{evtikhiev2023evaluating, hindle2016naturalness}.  

Traditional evaluation techniques rely on test-based methods such as pass@k, which assess code correctness by executing manually written test cases \cite{kulal2019spoc, chen2021evaluating}. While effective in certain contexts, these methods are limited by the need for extensive test case coverage, which is labor-intensive and may not fully capture code correctness beyond functional execution. Similarly, token-based metrics such as BLEU \cite{papineni2002bleu}, ROUGE-L \cite{lin2004rouge}, and CodeBLEU \cite{ren2020codebleu} have demonstrated weak correlations with human judgment when applied to code evaluation \cite{evtikhiev2023evaluating}.  

More recently, neural-based evaluation metrics such as CodeBERTScore \cite{zhou2023codebertscore} have shown improvements by leveraging deep learning models for code similarity assessments. However, these methods still depend on high-quality reference solutions, which can be expensive and difficult to obtain. Furthermore, even with neural models, code evaluation has not yet reached human-level understanding, as these models struggle to assess subtle correctness issues such as missing computation steps or failure to handle corner cases \cite{vaithilingam2022expectation, barke2022grounded}.  

Another key challenge is that AI-generated code is often more beneficial to experienced developers who can assess correctness, completeness, and security, whereas junior developers may struggle to critically evaluate and refine the generated code. Studies suggest that GitHub Copilot, one of the most widely used AI-powered coding assistants, provides significant advantages for senior developers but offers only limited value for junior developers \cite{arghavan2023github, vaithilingam2022expectation}.  

To address these limitations, we introduce a novel \textbf{reverse generation technique} combined with an \textbf{SBC score} to evaluate the accuracy and completeness of AI-generated code. Our approach does not rely on reference code but instead assesses how well the generated code aligns with the original requirement. This is achieved by extracting system requirements from the AI-generated code and comparing them with the initial requirements using three key metrics:  

\begin{itemize}
    \item \textbf{Semantic Similarity Score} – measuring meaning-level alignment between requirements,  
    \item \textbf{BLEU Score} – capturing lexical overlap,  
    \item \textbf{Completeness Score} – identifying missing and extra elements.  
\end{itemize}  

The SBC score, along with the reverse-generated requirements, provides actionable insights for developers, helping them assess AI-generated code without requiring extensive reference implementations. Unlike prior evaluation methods, this approach inherently addresses the challenges of syntactic variations and alternative solutions in generated code, as highlighted in recent studies \cite{tong2024syntactic}. By providing both quantitative scores and a human-readable requirement comparison, our method enhances transparency in AI-assisted software development, benefiting both junior and senior developers alike.

\section{Related Work}

There has been no prior work on applying a \textbf{reverse generation} technique with LLMs in conjunction with the \textbf{SBC scoring mechanism}. However, existing research has explored LLM-based evaluation methods for \textbf{code analysis} and \textbf{natural language generation (NLG)}, leveraging large models for assessing outputs. Three notable studies in this area include:

\subsection{G-EVAL: NLG Evaluation using GPT-4 with Better Human Alignment (Yang et al., \cite{yang2023geval})}

This paper introduces an \textbf{LLM-driven evaluation framework} that employs a \textbf{chain-of-thought (CoT) approach} to assess the quality of generated text. The study demonstrates that LLM-based evaluators can outperform traditional NLG metrics in \textbf{text summarization} and \textbf{dialogue generation}. However, the authors also highlight a critical limitation that LLM-based evaluators tend to exhibit a \textbf{bias toward LLM-generated text}, raising concerns about fairness and reliability in assessments.

\subsection{ICE-Score: Instructing Large Language Models to Evaluate Code (Terry, \cite{terry2024icescore})}

Inspired by \textbf{G-EVAL}, this paper proposes \textbf{ICE-Score}, an evaluation metric that leverages LLMs for code assessment across multiple programming languages, including \textit{Java, Python, C, C++, and JavaScript}. The approach incorporates both \textbf{human-centered usefulness} and \textbf{execution-based functional correctness}, aiming to provide a \textbf{holistic evaluation} of generated code. ICE-Score refines the instruction-based evaluation paradigm for assessing AI-generated code in a structured manner.

\subsection{Metamorphic Prompt Testing for LLM Code Validation (Xiaoyin et al., \cite{wang2024validating} )}

As Large Language Models (LLMs) become increasingly integrated into the software development lifecycle, concerns regarding the quality, correctness, and reliability of generated code have grown. Xiaoyin et al. \cite{wang2024validating} highlight these challenges and propose \textit{metamorphic prompt testing} as a validation approach. Their method leverages the intrinsic consistency among correct code samples while identifying inconsistencies in flawed ones. By paraphrasing prompts and generating multiple versions of code for cross-validation, their evaluation on HumanEval demonstrated that metamorphic prompt testing detected 75\% of erroneous programs generated by GPT-4, with a false positive rate of 8.6\%. This approach underscores the importance of rigorous validation techniques for ensuring the reliability of LLM-generated code in production environments.

This paper builds upon these advancements by extending the \textbf{LLM-driven evaluation paradigm} but introduces a \textbf{fundamentally different approach}. Rather than directly evaluating code output, we propose a \textbf{reverse generation} framework, where \textbf{requirements are inferred from code} and then compared to the original requirements using \textbf{semantic similarity, BLEU, and completeness} as core evaluation factors. This method not only aligns LLM-based evaluation with traditional NLP techniques but also provides \textbf{developers with a structured way to assess code accuracy and requirement fidelity}, bridging the gap between AI-generated code and human expectations.

\section{Methodology}

The following methodology was adopted in this study to evaluate the effectiveness of reverse generation and SBC scoring in AI-assisted development.

\subsection{Dataset Creation}

To construct a dataset that closely mimics the software development lifecycle, requirements were distributed across multiple layers of application development, such as:

\begin{itemize}
    \item \textbf{User Interface (UI)}: React, Angular
    \item \textbf{Data Layer}: SQL for data modeling
    \item \textbf{Data Objects and Business Logic}: .NET, Node.js, Quarkus, Spring Boot
\end{itemize}

A total of 90 requirements were curated, ensuring a diverse mix of technologies and application layers that reflect real-world developer usage. To prevent data contamination, as discussed by \cite{Aiyyappa2022} and \cite{terry2024icescore}, a completely new dataset was created instead of relying on pre-existing benchmark datasets.
The following figure illustrates few requirements from the dataset.
\vspace{-10pt}
\begin{figure}[H]
    \centering
    \includegraphics[width=0.9\textwidth]{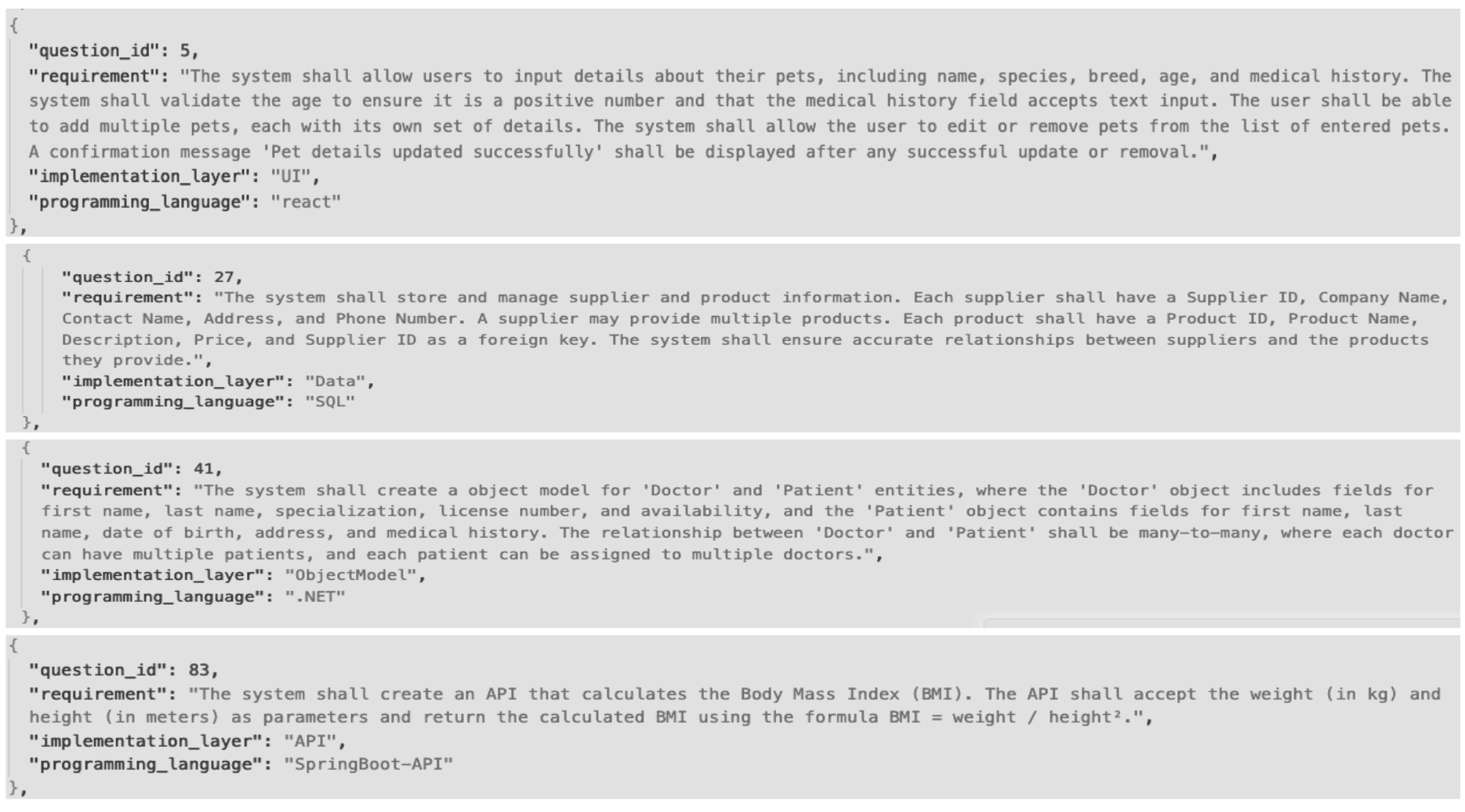}
    \caption{Requirements from various application layers in the dataset.}
    \label{fig:sample1}
\end{figure}

\subsection{Reference Implementation}

A reference implementation was developed in Python and is available in the associated \href{https://github.com/AhilanPonnusamy/Reverse-Generation-and-SBC-Metric--Reference-Implementation}{GitHub repository}. The implementation follows these key steps:

\begin{enumerate}
    \item Iterate through the requirements and target technologies in the dataset.
    \item Invoke the LLM to generate code for the given requirement.
    \item Perform reverse generation by passing the generated code back to the LLM with a detailed prompt to reconstruct the requirement.
    \item Compare the original requirement with the reverse-generated requirement using the SBC scoring mechanism.
    \item Store results in JSON format, including input and generated requirements, final SBC score, and individual component scores (semantic similarity, BLEU, completeness).
\end{enumerate}

\subsection{Choice of Open Models}

The existing studies highlighted in the Related Work section \cite{yang2023geval, terry2024icescore} use closed-source hosted models, such as GPT-3.5 and GPT-4, for their analysis. In contrast, this study focuses on open models, as discussed in \cite{white2024mof}. 

We selected four quantized models (Q4 or Q5) for this study:

\begin{itemize}
    \item \textbf{Model 1:} Codellama 13B
    \item \textbf{Model 2:} Qwen2.5-Coder 14B
    \item \textbf{Model 3:} Deepseek Coder 6.7B
    \item \textbf{Model 4:} Codestral 22B
\end{itemize}

These mid-level models were chosen for their balance between size, efficiency, and capability in handling both code and natural language tasks effectively.

\subsection{SBC Score Calculation}

The SBC score is computed using the formula:

\begin{equation}
\text{fs} = (0.7 \times \text{semantic\_score}) + (0.1 \times \text{BLEU}) + (0.2 \times \text{completeness})
\end{equation}

where:

\begin{itemize}
    \item \textbf{Semantic Similarity} is computed using the PyTorch \texttt{cos\_sim} function to measure the cosine similarity between input and generated requirement encodings using the \texttt{all-MiniLM-L6-v2} model from the Sentence Transformers library.
    \item \textbf{BLEU Score} evaluates n-gram matching between the original and reverse-generated requirements.
    \item \textbf{Completeness Score} extracts key nouns, verbs, and proper nouns from the input and computes a penalty based on missing or extra keywords:
\end{itemize}

\begin{equation}
\text{keywords1} = \text{extract\_keywords}(\text{input requirements from dataset})
\end{equation}
\begin{equation}
\text{keywords2} = \text{extract\_keywords}(\text{reverse generated requirements})
\end{equation}
\begin{equation}
\text{missing} = \text{keywords1} - \text{keywords2}
\end{equation}
\begin{equation}
\text{extra} = \text{keywords2} - \text{keywords1}
\end{equation}
\begin{equation}
\text{total\_keywords} = \left|\text{keywords1} \cup \text{keywords2}\right|
\end{equation}
\begin{equation}
\text{penalty} = \left|\text{missing}\right| + \left|\text{extra}\right|
\end{equation}
\begin{equation}
\text{score} = \max\left( 0, 1 - \frac{\text{penalty}}{\max\left(\text{total\_keywords}, 1\right)} \right)
\end{equation}

\vspace{\baselineskip}
The weight distribution in the SBC score reflects the relative importance of each metric in capturing requirement-code alignment. 
\textbf{Semantic similarity} is given the highest weight (0.7) as it directly measures how closely the reverse-generated requirement aligns with the original intent, making it the most critical factor. 
\textbf{Completeness} is weighted at 0.2 to ensure that missing or extra keywords are penalized without dominating the score, while \textbf{BLEU}, often less reliable for evaluating long-form text generation, is assigned a lower weight (0.1) to complement semantic similarity without disproportionately influencing the final score.

\subsection{A Note on Correlation Metrics}

While correlation coefficients such as Kendall-Tau ($\tau$), Pearson ($r_p$), and Spearman ($r_s$) are commonly used to measure relationships between variables, they were not included in the SBC score computation. These metrics primarily assess agreement and ranking consistency, whereas SBC is designed as a weighted function of semantic similarity, BLEU, and completeness, to directly measure requirement-code alignment.

That said, these correlation measures could be valuable for post-hoc validation, particularly in assessing how well SBC correlates with human judgment or other evaluation metrics. Future research may explore their role in further validating SBC’s effectiveness in AI-assisted development scenarios.

\subsection{Visualization and Analysis}

For visualization and analysis, this study utilized Google Sheets to generate graphs and insights from the SBC score data. The process involved:

\begin{itemize}
    \item Exporting the SBC score output into a CSV format.
    \item Creating a Pivot Table to aggregate and structure the data.
    \item Generating graphs and charts to visualize trends in the SBC scores.
\end{itemize}

The primary focus was on the \textbf{final SBC score}, as the \textbf{semantic similarity component} had a disproportionately strong influence on the overall score compared to BLEU and completeness. The generated graphs provided valuable insights into trends and comparative performance across different models and technologies.

\section{Experiments and Results}

To ensure consistency in responses, we conducted all tests with the temperature set to zero. With the reference implementation, we conducted the evaluation in two cycles:  

\subsection{Per-LLM Evaluation}  
Each LLM was run for three iterations across all 90 questions in the dataset. The SBC score details were recorded in JSON format, capturing key elements such as \texttt{input\_requirements}, \texttt{reverse\_generated\_requirements}, \texttt{final\_accuracy\_score}, \texttt{semantic\_similarity}, \texttt{BLEU\_score}, \texttt{completeness\_score}, \texttt{missing\_elements}, and \texttt{extra\_elements}.  The figure below presents the evaluation graph for each LLM included in this study, providing a comparative analysis of their performance.

\vspace{-10pt}
\begin{figure}[H]
    \centering
    \includegraphics[width=0.9\textwidth]{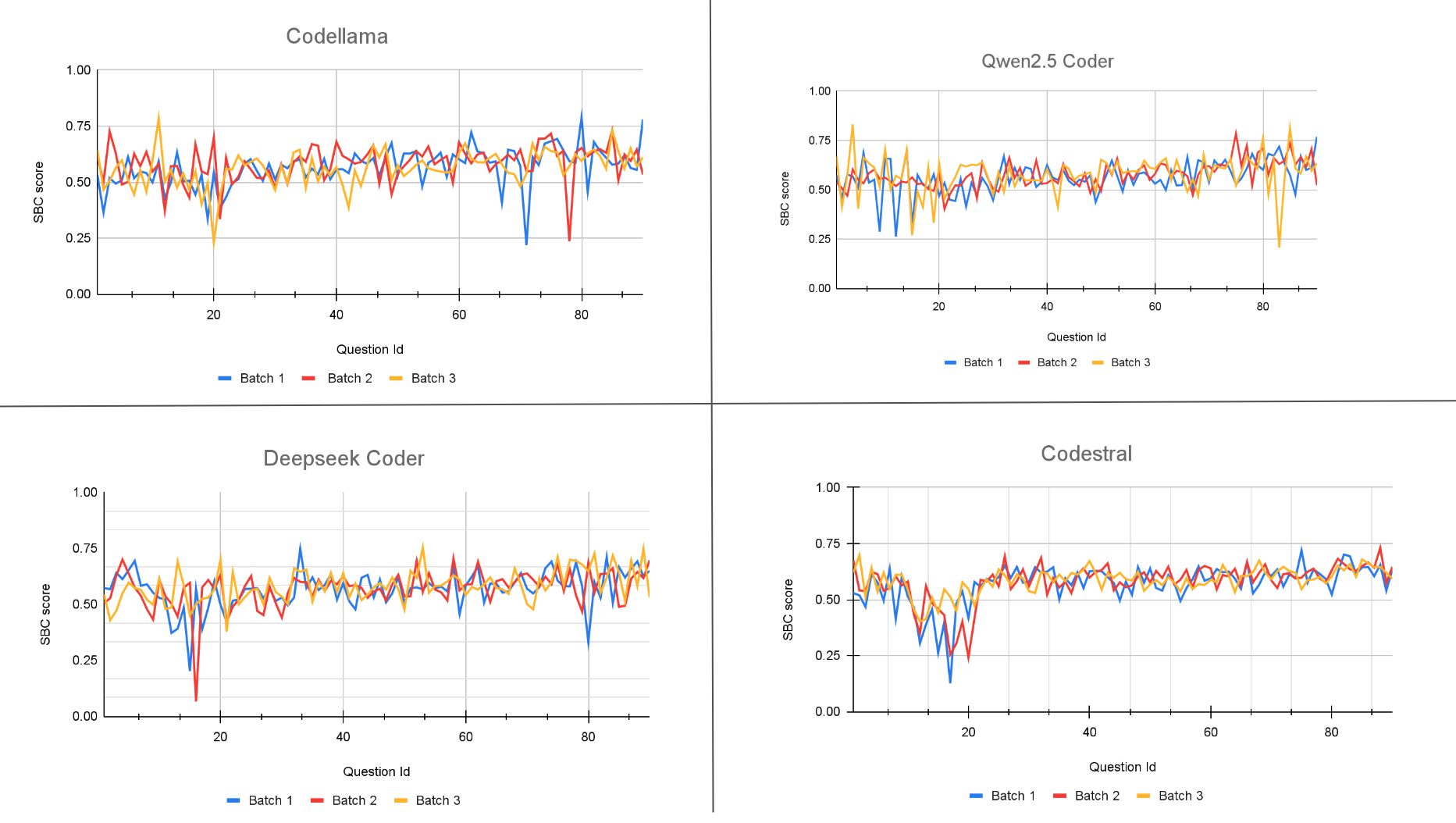}
    \caption{Performance evaluation graph for all LLMs over three iterations.}
    \label{fig:sample2}
\end{figure}

Notably, the \texttt{missing\_elements} section provides valuable insight into the requirement components that were omitted during code generation, while the \texttt{extra\_elements} section serves as a strong indicator of hallucinations introduced by the LLM. These insights help developers identify potential gaps in the LLM-generated code and deviations, including hallucinations. The results from these iterations were then plotted and analyzed for comparison.  The following figure illustrates a generated SBC response from the experiment.

\vspace{-10pt}
\begin{figure}[H]
    \centering
    \includegraphics[width=0.9\textwidth]{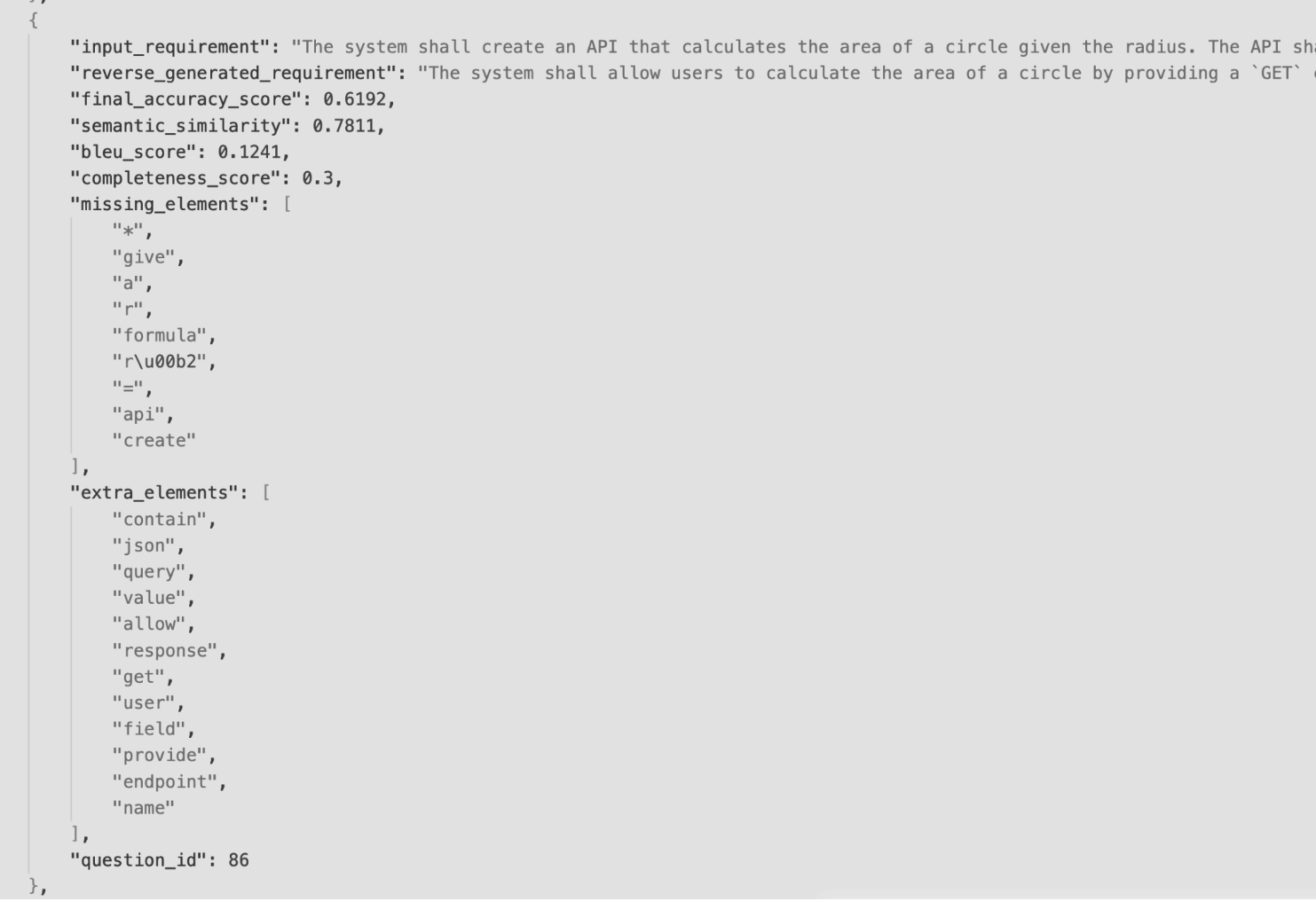}
    \caption{Generated response with SBC score for Question 86 from the Deepseek Coder model.}
    \label{fig:sample3}
\end{figure}

\subsection{Cross-LLM Comparison}  
After evaluating all four LLMs, resulting in a total of 12 iterations, the results were consolidated into a final comparative analysis. For this, the maximum SBC score for each question within each LLM was extracted to form a representative dataset. A final consolidated chart was then created to compare the performance of all LLMs based on these maximum scores.  

The results show that the line graphs for all LLMs are closely aligned, indicating that their performance trends are similar, rising and falling together across different questions. Additionally, we observed that across technologies and application layer questions, the LLMs performed consistently with minimal variance.  

The following figure illustrates the consolidated graph below presents the maximum SBC scores for each LLM across the three iterations.  

\vspace{-10pt}
\begin{figure}[H]
    \centering
    \includegraphics[width=0.9\textwidth]{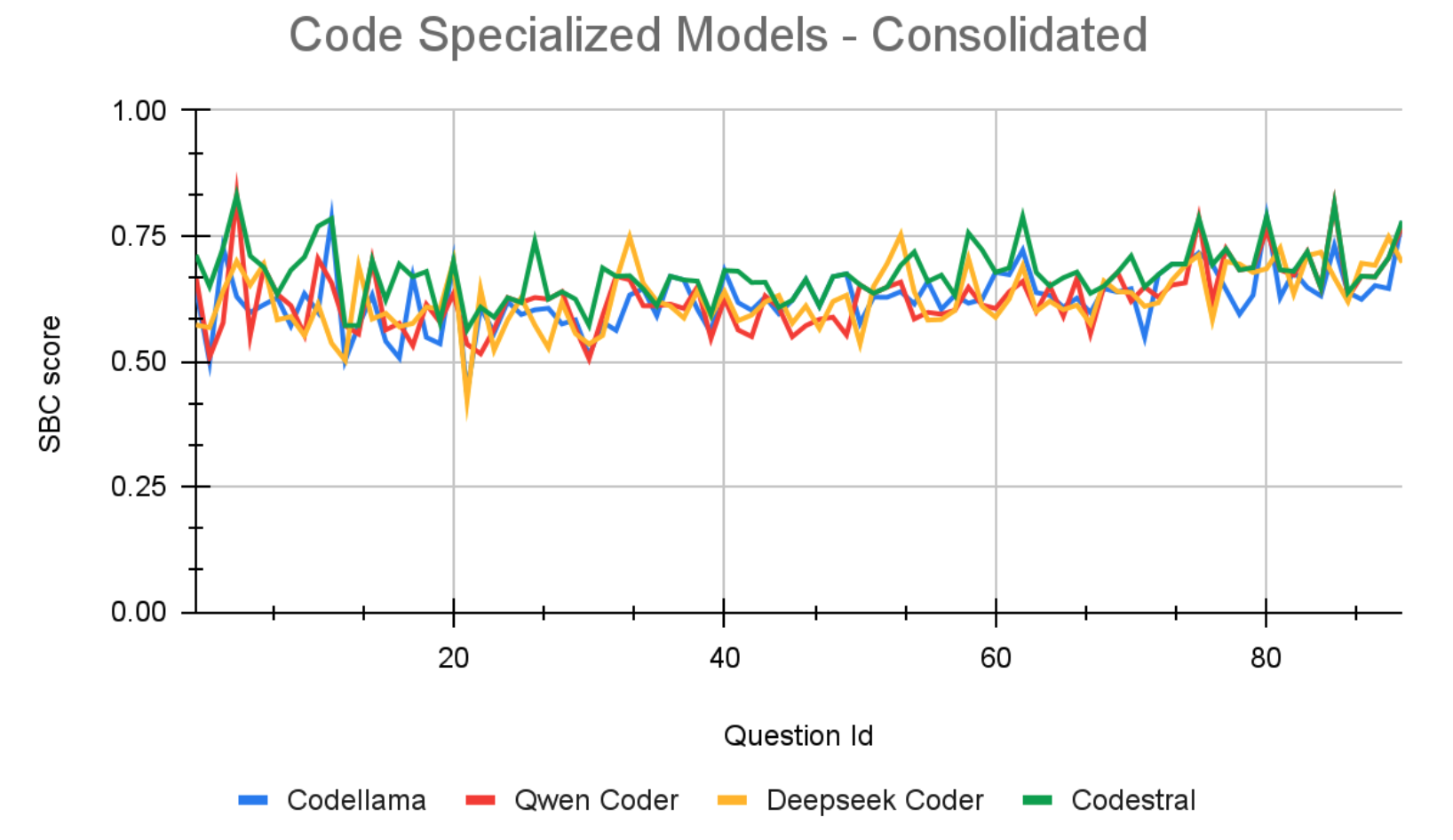}
    \caption{Consolidated score with max scores  for all LLMs.}
    \label{fig:sample4}
\end{figure}

We observed that reverse-generated requirements are of high quality and easy to interpret when SBC scores exceed 0.55 as shown in the following figure. 

\vspace{-10pt}
\begin{figure}[H]
    \centering
    \includegraphics[width=0.9\textwidth]{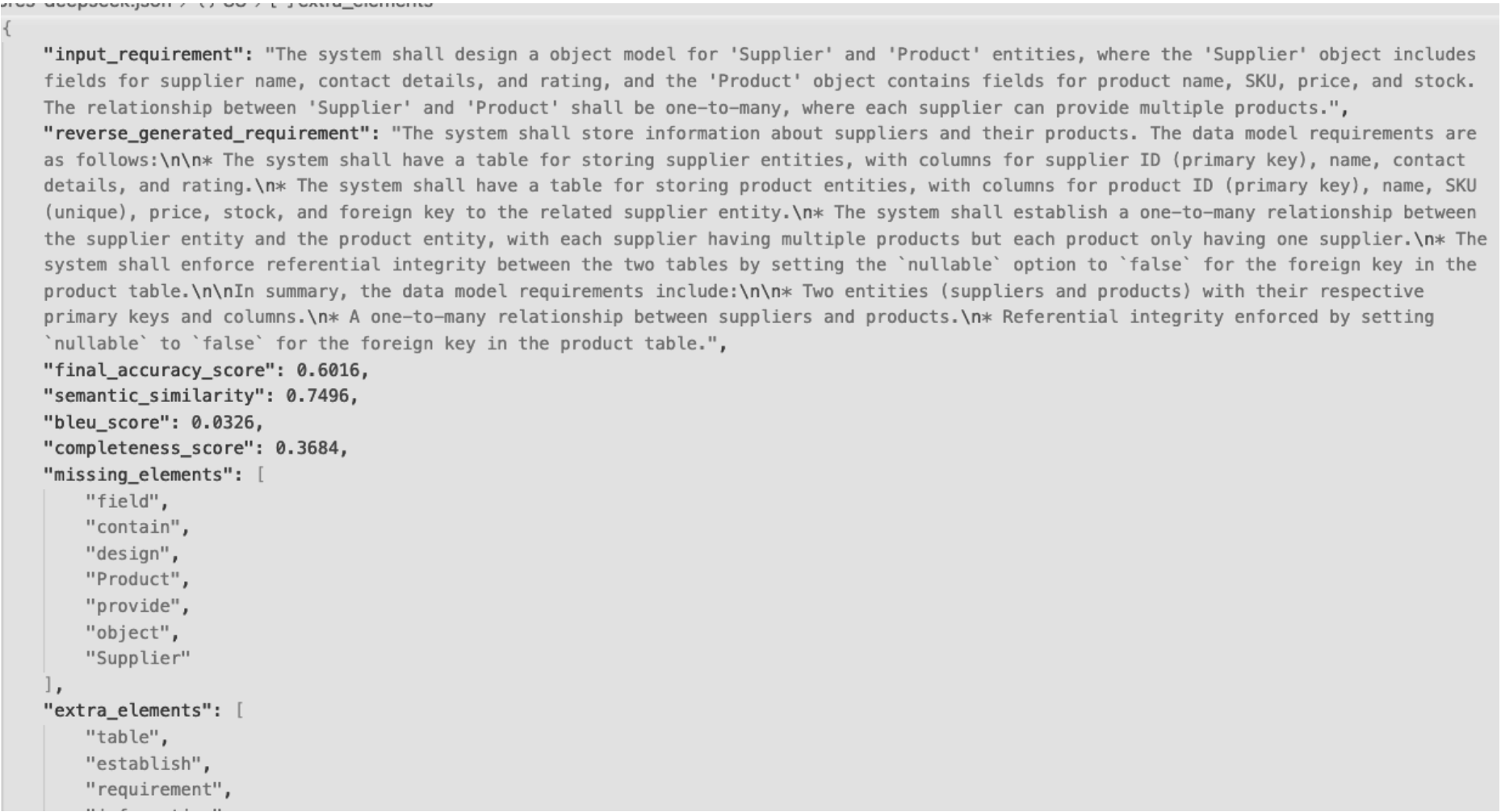}
    \caption{ Sample output for SBC score above 0.55.}
    \label{fig:sample5}
\end{figure}

For SBC scores above 0.65, the generated requirements were found to be semantically very close to the original inputs as shown in the following figure. 

\vspace{-10pt}
\begin{figure}[H]
    \centering
    \includegraphics[width=0.9\textwidth]{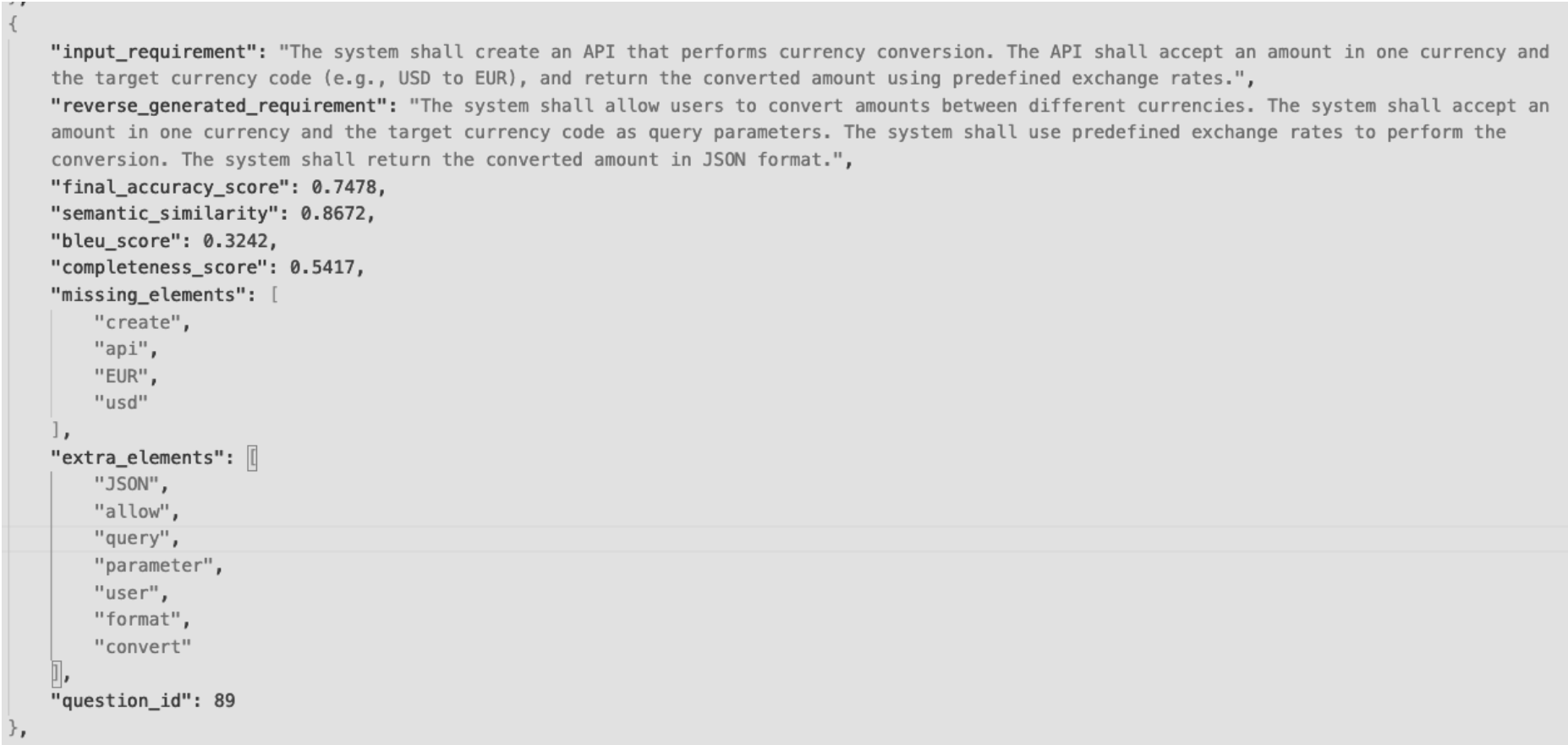}
    \caption{ Semantically similar output with SBC score above 0.65.}
    \label{fig:sample6}
\end{figure}

\subsection{Integrating SBC Score and Artifacts with the Application Development Life Cycle}

Given the significant benefits that the SBC score, along with missing and extra keyword analysis, can provide, it is highly beneficial for enterprises to integrate this process into the AI code assistant workflow. Instead of merely returning the generated code, AI code assistant responses can be modified  to include the SBC score along with reverse-generated requirements, and highlight missing and extra keywords. This integration empowers developers at all levels, ensuring that the advantages of AI-driven code generation are accessible to a broader range of team members. By providing these insights, entrprises can better identify and mitigate potential issues and risks at an earlier stage, improving the overall efficiency and quality of the software development life cycle.

\section{Conclusion and Next Steps}  
In this study, we introduced a novel approach to evaluating LLM-generated code by leveraging reverse generation and the Semantic-BLEU-Completeness (SBC) hybrid metric. Our methodology provides a tool to validate if the  code generation aligns closely with initial requirements by measuring semantic similarity, BLEU score, and completeness. The structured evaluation across four LLMs, conducted with a zero-temperature setting for consistency, demonstrated that most models performed similarly, with their performance trends rising and falling in a closely knitted pattern.  

A key objective of this work is to provide a solution that serves both senior and junior developers by enabling them to assess the validity and completeness of LLM-generated code without the need for manual inspection of the code itself. Our approach allows developers to quickly evaluate whether the generated code aligns with the original requirements, making it accessible and useful to professionals with varying levels of expertise. This minimizes the time and effort needed to validate code, empowering developers to focus on more complex tasks and improving overall productivity.  

Our results highlight the strengths and limitations of LLMs in generating reliable code from natural language requirements. The analysis of \texttt{missing\_elements} and \texttt{extra\_elements} provided valuable insights into common failure modes, revealing both missing functionalities and hallucinations in generated code. These insights suggest that reverse generation can serve as an effective validation mechanism, enabling developers to assess and refine LLM-generated code systematically.  

Unlike prior studies that predominantly leveraged closed hosted cloud models such as GPT-3.5 and GPT-4 \cite{terry2024icescore, yang2023geval}, we specifically chose open models to ensure transparency, reproducibility, and fine-grained control over the deployment and the evaluation process. This decision aligns with our goal of establishing a more flexible, interpretable, and adaptable framework for assessing LLM-generated code.

Despite the promising results, several areas warrant further exploration. While we relied on SBC scores to compare alignment between input requirements and generated code, human feedback remains an essential validation mechanism. Future work should incorporate qualitative assessments from software engineers to further validate whether high SBC scores correlate with human perceived correctness and usability of the generated code. Additionally, while correlation coefficients such as Pearson ($r_p$), Spearman ($r_s$), and Kendall-Tau ($\tau$) were not included in SBC score computation, they remain useful validation tools for analyzing how well SBC scores align with human evaluations. Future experiments can leverage these correlation measures to assess whether the SBC metric effectively captures human judgment of requirement-code alignment.  

Another direction for future research involves expanding the dataset to cover a wider range of programming languages and problem domains, including more complex algorithmic tasks and API-driven implementations. Moreover, investigating the impact of prompt engineering techniques and fine-tuning open model LLMs for code generation could further enhance alignment with intended requirements especially in enterprise context.

By bridging the gap between LLM-generated code and human defined requirements, our approach provides a structured and interpretable framework for evaluating AI-assisted development. As LLMs continue to evolve, refining hybrid metrics such as SBC and integrating human feedback will be crucial to ensuring their effectiveness in real-world software engineering workflows.

\bibliographystyle{ieeetr}  % Use plain bibliography style
\bibliography{references}  % Ensure you have a references.bib file

\end{document}